\documentclass[prb,preprint,showpacs]{revtex4}% Physical Review B

\usepackage{bbm}
\usepackage{epsfig}
\begin{document}
\title{Magnetoswitching of current oscillations in diluted magnetic semiconductor 
nanostructures}
\author{R. Escobedo$^1$, M. Carretero$^{2,3}$, L. L. Bonilla$^{2,3}$ and G. 
Platero$^4$}
%\address{%
\affiliation {
$^1$Departamento de Matem\'atica Aplicada y Ciencias de la Computaci\'on,
Universidad de Cantabria, 39005 Santander, Spain\\
$^2$G. Mill\'an Institute, Fluid Dynamics, Nanoscience and Industrial Mathematics,
Universidad Carlos III de Madrid, 28911 Legan\'es, Spain\\
$^3$Unidad Asociada al Instituto de Ciencia de Materiales de Madrid, CSIC, Spain\\
$^4$Instituto de Ciencia de Materiales de Madrid, CSIC, 28049 Cantoblanco, Madrid, Spain}
%\ead{escobedo@unican.es, manuel.carretero@uc3m.es, bonilla@ing.uc3m.es and gplatero@icmm.csic.es}
\date{\today}
%%%%%%%%%%%%%%%%%%%%%%%%%%%%%%%%%%%%%%%%%%%%%%%%%%%%%%%%%%%%%%%%%%
%%%%%%%%%%%%
\begin{abstract}
Strongly  nonlinear transport  through Diluted Magnetic Semiconductor 
%(DMS) 
multiquantum wells occurs due to the interplay between confinement, Coulomb and exchange 
interaction. Nonlinear effects include the appearance of spin polarized stationary states and 
self-sustained current oscillations as possible stable states of the nanostructure, depending on 
its configuration and control parameters such as voltage bias and level splitting due to an 
external magnetic field. Oscillatory regions grow in size with well number and level splitting.
A systematic analysis of the charge and spin response to voltage and magnetic field switching 
of II-VI Diluted Magnetic Semiconductor multiquantum wells is carried out. The description 
of stationary and 
time-periodic spin polarized states, the transitions between them and the responses to voltage 
or magnetic field switching have great importance due to the potential implementation of 
spintronic devices based on these nanostructures.
%The response to voltage or magnetic field switching of a n-doped dc 
%voltage biased II-VI semiconductor multiquantum well structure having its first well doped 
%with magnetic impurities (Mn) is analyzed. 
%The phase diagram of voltage versus level splitting caused by an external magnetic field 
%shows regions of stable self-sustained current oscillations immersed in others of stable 
%stationary states. Oscillatory regions grow in size with well number and level 
%splitting. Sudden voltage or magnetic field changes may switch or disconnect current 
%oscillations from an initial stationary state. 
\end{abstract}
%%%%%%%%%%%%%%%%%%%%%%%%%%%%%%%%%%%%%%%%%%%%%%%%%%%%%%%%%%%%%%%%%%
%%%%%%%%%%%%
%\begin{multicols}{2}
\pacs{85.75.-d, 72.25.Dc, 75.50.Pp, 73.63.Hs}

\maketitle

\section{Introduction}
Spin injection is one of the aims of spintronics \cite{spi} thanks to the potential 
applications of injectors as spin LED devices, etc. Also quantum state transfer from 
spin electrons to photons by interband  transitions is actively investigated \cite{ima,cer05,titov}.
One of the most efficient ways of spin injection to date \cite{fie,sch} is the use of II-VI 
dilute magnetic semiconductors (DMSs) that exhibit the giant Zeeman effect \cite{fur}: 
they have a conductivity comparable to that of nonmagnetic semiconductors, and can boast 
spin polarizations close to 100\% at a small applied magnetic field. However, spin-injection 
experiments in semiconductors enter easily the regime of nonlinear  response \cite{david2}. 
Different effects could contribute to nonlinear transport and therefore to nonlinear spin 
injection. For example, band bending effects \cite{sch1} in nanostructures give rise to 
a nonlinear current due to the interplay between Coulomb interaction  and electron tunnel in 
these confined systems, which have quasi-discrete states.

Other physical mechanisms inherent to these systems explain their current-voltage 
characteristics: for instance a large Zeeman level splitting $\Delta$ in an applied magnetic 
field $B$. Recently, spin transport through DMS diodes \cite{fabian} and
multi-quantum well structures (MQWS) has been analyzed \cite{david,bejar,apl,njp}. These 
works study nonlinear features of the current (hysteresis, multistability) as a function of the 
external voltage. Under strong dc voltage bias $V$, electric field domains are formed in 
MQWS due to the interplay between electron-electron interaction and resonant tunneling
\cite{david}. In other sample configurations or, for different doping density, there are spin
polarized self-sustained current oscillations (SSCOs) and the system could behave as a spin
oscillator \cite{bejar,apl}. To tailor the properties of these spin oscillators or injectors,
it is important to perform a systematic analysis of the transition from stationary to time
dependent current, in terms of sample configuration, external magnetic field, doping density,
etc.

In this paper we analyze the response to voltage ($V$) or magnetic ($B$) switching in a
n-doped dc voltage biased semiconductor MQWS having its first quantum well (QW) doped with Mn.
Both spin polarized stationary states (SSs) and SSCOs are possible stable states of the MQWS
for different values of the parameters. Stationary states field profiles consist of two
electric field domains separated by a domain wall which is a charge monopole \cite{njp}.
Magnetic field switching requires knowing phase diagrams of the current density $J$ and the
applied voltage $V$ versus the level splitting $\Delta$ (due to the magnetic field $B$), and
these diagrams are among the results of this paper.
The phase diagram of $V$ versus $\Delta$ shows regions of stable SSCOs embedded in 
others of stable SSs. The extension of the SSCO regions increases with the number of QWs 
in the structure. Sudden changes of $V$ or $B$ may switch or disconnect SSCOs from an 
initial stable SS or force the domain wall to change its location. The SSCOs are due 
to periodic triggering of charge dipoles at the Mn-doped well and their motion towards the 
collector \cite{apl}. Large level splitting induced by $B$ due to the exchange interaction 
provides DMS MQWSs with a new degree of freedom which is absent in conventional III-V
 weakly coupled n-doped semiconductor MQWSs \cite{BGr05}. Another important difference is
 that, in the latter, both charge dipoles and monopoles may be triggered at the
injector (depending on its current--field characteristics: its conductivity if the relation between
current and field is linear) and both may cause SSCOs \cite{san99,BGr05}. In 
these materials and for moderate conductivity of the injecting contact, switching the voltage 
$V$ between different SSs involves either upward
monopole motion or a dipole-tripole mechanism \cite{amann,bon06,BGr05}. For sufficiently
large conductivity of the injector, the dipole-tripole mechanism ceases to exist and voltage
switching involves injection of a charge monopole that moves towards the collector until it
reaches the QW corresponding to the final stable SS \cite{hu07}. Voltage or magnetic
switching in II-VI MQWSs always involves dipole nucleation at the Mn-doped QW. 

\section{Model}
Our sample configuration consists of an n-doped ZnSe/(Zn,Cd,Mn)Se weakly coupled
MQWS. The spin for the magnetic ion $Mn^{++}$ is $S$=5/2 and the exchange interaction
between the Mn local moments and the conduction band electrons is ferromagnetic in II-VI
QWs. The energy spectrum corresponding to $N$ isolated QWs comprising our weakly coupled MQWS
has the form $E_j + \hbar^2k_{\perp}/(2m^*)$, where $m^*$ is the effective mass, $k_\perp$ is the 
in-plane wave vector orthogonal to the growth direction and $j=1,\ldots, N$ is the QW subband index. 
In the weak magnetic fields considered here, we disregard Landau-level formation and $k_\perp$ is 
a continuous variable \cite{david}. Using the virtual crystal and mean field approximations, the exchange
interaction causes the subband energies to depend on spin in those QWs containing Mn ions:
\begin{eqnarray}
E_j^{\pm}=E_j \mp \Delta/2,
\end{eqnarray}
 where 
\begin{eqnarray}
 \Delta (B) = 2 J_{\rm sd} N_{\rm Mn} S B_S
 \left( {g \mu_B S \over k_B T_{\rm eff}} B \right)
\label{d.b}
\end{eqnarray}
for spin $s=\pm 1/2$, and $B_S$, $J_{sd}$, $N_{Mn}$ and $T_{\rm eff}$ are the 
Brillouin function, the exchange integral, the density of magnetic impurities and
an effective temperature which accounts for Mn interactions, respectively \cite{david,slobo}.

We model spin-flip scattering coming from spin-orbit or hyperfine interaction by a 
phenomenological scattering time $\tau_{\rm sf}$, which is larger than impurity and phonon 
scattering times: $\tau_{\rm scat}<\tau_{\rm sf}$. Vertical transport in the weakly coupled MQWS is 
spin-independent sequential tunneling between adjacent QWs, so that when electrons tunnel to 
an excited state they instantaneously relax by phonon scattering to the ground state, with the 
same spin polarization \cite{david}. Lastly, electron-electron interaction is considered within the 
Hartree mean field approximation.

The equations governing the model are \cite{apl,njp}: the discrete
Poisson equation relating the  two-dimensional spin-up and spin-down electron densities, 
$n^+_i$, $n_{i}^-$, respectively, to the average electric field $-F_{i}$ at the $i$th MQWS period 
(of lengh $l$),
\begin{eqnarray}
\varepsilon\, (F_i - F_{i-1}) =  e\, (n_i^+ + n^-_{i} - N_D),
\label{e1}
\end{eqnarray}
and the rate equations for $n^\pm_{i}$,
\begin{eqnarray}
e\, \frac{{\rm d}n^\pm_i}{{\rm d}t} =
J^\pm_{i-1\rightarrow i} - J^\pm_{i \rightarrow i+1}
\pm \frac{A(n_i^+,n_i^-,\mu_{i}^+)}{\tau_{\rm sf,i}},
\label{e2}
\end{eqnarray}
for $i=1,\dots, N$. For numerical convenience, we have introduced here a smoothed form
$A(n_i^+,n_i^-,\mu_{i}^+)$ of the scattering term used in~\cite{david}, given by
\begin{eqnarray}
A(n_i^+,n_i^-,\mu_{i}^+) = n_{i}^-
 - \frac{n_{i}^+}{1 + \exp \left( \frac{E_{1,i}^- - \mu_{i}^+}{\gamma_{\mu}} \right)},
\end{eqnarray}
where $\gamma_{\mu}$ is a small smoothing parameter (smaller than $\gamma=\hbar/\tau_{
\rm scatt}$ or than the thermal energy) such that, as $\gamma_{\mu}\to 0$,
%$\pm A(n_i^+,n_i^-,\mu_{i}^+)$ becomes $\pm(n^-_{i}-n^+_{i})$ for
%$\mu_{i}^+ > E_{1,i}^-$ (equivalently, $\mu_{i}^+ - E_{1,i}^+ > \Delta$),
%and $\pm n^-_{i}$ otherwise \cite{david}.
%
\begin{eqnarray*}
 \pm A(n_i^+,n_i^-,\mu_{i}^+) \to \pm n^-_{i} + \left\{
\begin{array}{ll}
 \mp n^+_{i} & \mbox{ for } \mu_{i}^+ > E_{1,i}^-, \\
 0 & \mbox{ otherwise,}
\end{array}
\right.
\end{eqnarray*}
which was used by S\'anchez et al \cite{david} 
(note that $\mu_{i}^+ > E_{1,i}^-$ is equivalent to $\mu_{i}^+ - E_{1,i}^+ > \Delta$).

In these expressions, $\mu_i^\pm$ is the chemical potential at the $i$th MQWS 
period and $E_{j,i}^\pm$ are the spin-dependent subband energies (measured from the
bottom of the $i$th well): $E_{j,1}^\pm = E_{j}\mp \Delta/2$, and $E_{j,i}^\pm$=$E_{j}$ 
for $i\neq$1. Also, $N_D$ and $\varepsilon$ are the 2D doping density at the
QWs and the average permittivity.

In weakly coupled MQWS, tunneling between adjacent QWs can be treated in 
leading order perturbation theory. Since elastic
and inelastic scattering times in the QWs are
shorter than any other time scale of the problem, we can assume that
the electrons in each well are in quasiequilibrium between
succesive tunneling events and that their temperature is that of the lattice.
We ignore interwell spin-flip processes, so that
currents are carried between wells by the two spin subsystems
in parallel. Then, as in the case of non-magnetic MQWSs, the tunneling current densities 
across the $i$th barrier $J_{i\to i+1}^\pm$ can be calculated by the Bardeen Transfer Hamiltonian 
method \cite{bardeen,BPT88a,BPT88b,BPT89}. See the detailed derivation for 
non-magnetic MQWSs in Ref.~\cite{hu07}. The well known resulting expression 
\cite{hu07,Bon02} can be approximated by the formula \cite{njp}:
\begin{eqnarray}
J_{i\to i+1}^\pm & = & {e\, v^{(f)\pm} (F_i) \over l} \left\{n_i^\pm - {m^* k_B T
\over 2\pi\hbar^2} \ln \left[ 1 + e^{ - {e F_i l \over k_B T}}  \right.
\right.   \nonumber \\
&& \left. \left. \times
\left( \exp \left({2\pi \hbar^2 n_{i+1}^\pm \over m^* k_B T} \right) -1
 \right)\right] \right\}, \label{e5}
\end{eqnarray}
$i=1,\dots,N-1$, provided that scattering-induced broadening of energy levels is much
smaller than subband energies and chemical potentials; see Appendix A of Ref.~\cite{Bon02}. 
The spin-dependent  ``forward tunneling velocity'', $v^{(f)\pm}$, is a sum of Lorentzians of width
$2\gamma$, with $\gamma=\hbar/\tau_{\rm scatt}$ (the same value for all subbands, for 
simplicity), centered at the resonant field values
$F^\pm_{j,i}= (E^\pm_{j,i+1}- E^\pm_{1,i})/(el)$,
\begin{eqnarray}
v^{(f)\pm}(F_i)= \frac{\hbar^3 l \gamma}{2\pi^2 m^{* 2}}\sum_{j=1}^2 \frac{\mathcal{T}_i(E^\pm_{1,i})
}{ (F_i - F^\pm_{j,i})^2 (e l)^2+(2\gamma)^2},\label{e55}
\end{eqnarray}
where $\mathcal{T}_i$ is proportional to the transmission coefficient of the $i$th barrier
\cite{Bon02}. For electrons with spin $\pm 1/2$, the chemical potential $\mu_{i}^\pm$ and the 
electron densities $n_{i}^ \pm$ are related by
\begin{eqnarray}
n_{i}^\pm = \frac{m^*k_B T}{2\pi \hbar^2} \,
\ln \left[ 1 + \exp \left( \frac{\mu_i^\pm - E_{1,i}^\pm}{k_{B}T} \right)  \right].
\end{eqnarray}

The voltage bias condition can be written as 
\begin{eqnarray}
\sum_{i=0}^N F_i l = V.
\end{eqnarray}

Defining $J_{i\rightarrow i+1} = J_{i\rightarrow i+1}^+ +J_{i\rightarrow i+1}^-$,
the total current density $J(t)$ can be calculated as
\begin{eqnarray}
J(t)=\frac{1}{N+1}\,\sum_{i=0}^N J_{i\rightarrow i+1}.
\label{e3}
\end{eqnarray}
Then, time-differencing the Poisson equation, inserting the rate equations for
$n^\pm_{i}$ in the result, and assuming a constant applied voltage ($dV/dt=0$),
we obtain the following equation relating $F_i(t)$, $J_{i\rightarrow i+1}(t)$
and $J(t)$ for $i=0,\dots, N$:
\begin{eqnarray}
\varepsilon\, \frac{{\rm d}F_i}{{\rm d}t} + J_{i\rightarrow i+1} = J(t).
\label{e3b}
\end{eqnarray}

Boundary tunneling currents for $i=0$ and $i=N$ are determined by using tunneling
currents with $n^\pm_{0}=n_{N+1}^\pm= N_{D}/2$ (identical normal contacts) \cite{david}.

As initial conditions, we set $n_i^\pm=N_{D}/2$ (normal QWs) and $F_{i}=\phi F_{M}$,
where $F_{M}$ is a reference field corresponding to the first local maximum $(F_M,J_M)$ of
the tunneling current $J_{i\rightarrow i+1}(F,n_i^+,n_i^-,n_{i+1}^+,n_{i+1}^-)$ for 
$n_i^{\pm} = n_{i+1}^{\pm} =  N_D/2$ in a nonmagnetic well \cite{njp}, and $\phi$ is a 
dimensionless average field defined by
\begin{eqnarray}
\phi = \frac{V}{l (N+1) F_M}.
\end{eqnarray}
$(N+1)\phi$ is the dimensionless voltage across the MQWS.

\section{Results}
We have considered barrier and QW widths of 10 and 5 nm, respectively, 
$\tau_{\rm sf}$=$10^{-9}$s
(normal QW) and $10^{-11}$s (magnetic QW), $m^*$=$0.16 m_{0}$, 
$N_{D}$=$10^{10}$cm$^{-2}$, $\varepsilon$=$7.1 \varepsilon_{0}$, $T$=$5$ K, 
$E_{1}$=$15.76$ meV, $E_2$=$61.99$ meV, $\gamma$=$1$ meV and 
$\gamma_{\mu}$=$0.1$ meV  \cite{apl}. 

To find the relation between $\Delta$
and $B$, we used the values $g=2$, $S=5/2$ and $T_{\rm eff}=T+T_0$ with $T_0=2$ K.
The prefactor in (\ref{d.b}) can be estimated from Fig.\ 3 of Ref.\ \cite{slobo} to be 23.26 meV.
Then, for $T=5$ K, we find 
\begin{eqnarray}
 \Delta (B) = 23.26\, B_{5/2}( B/ 2.084);
\label{ZS}
\end{eqnarray}
units of $B$ and $\Delta$ are Tesla and meV, respectively.

\begin{figure}[ht]
\centerline{\hbox{
\epsfxsize=.85\textwidth
\epsfbox{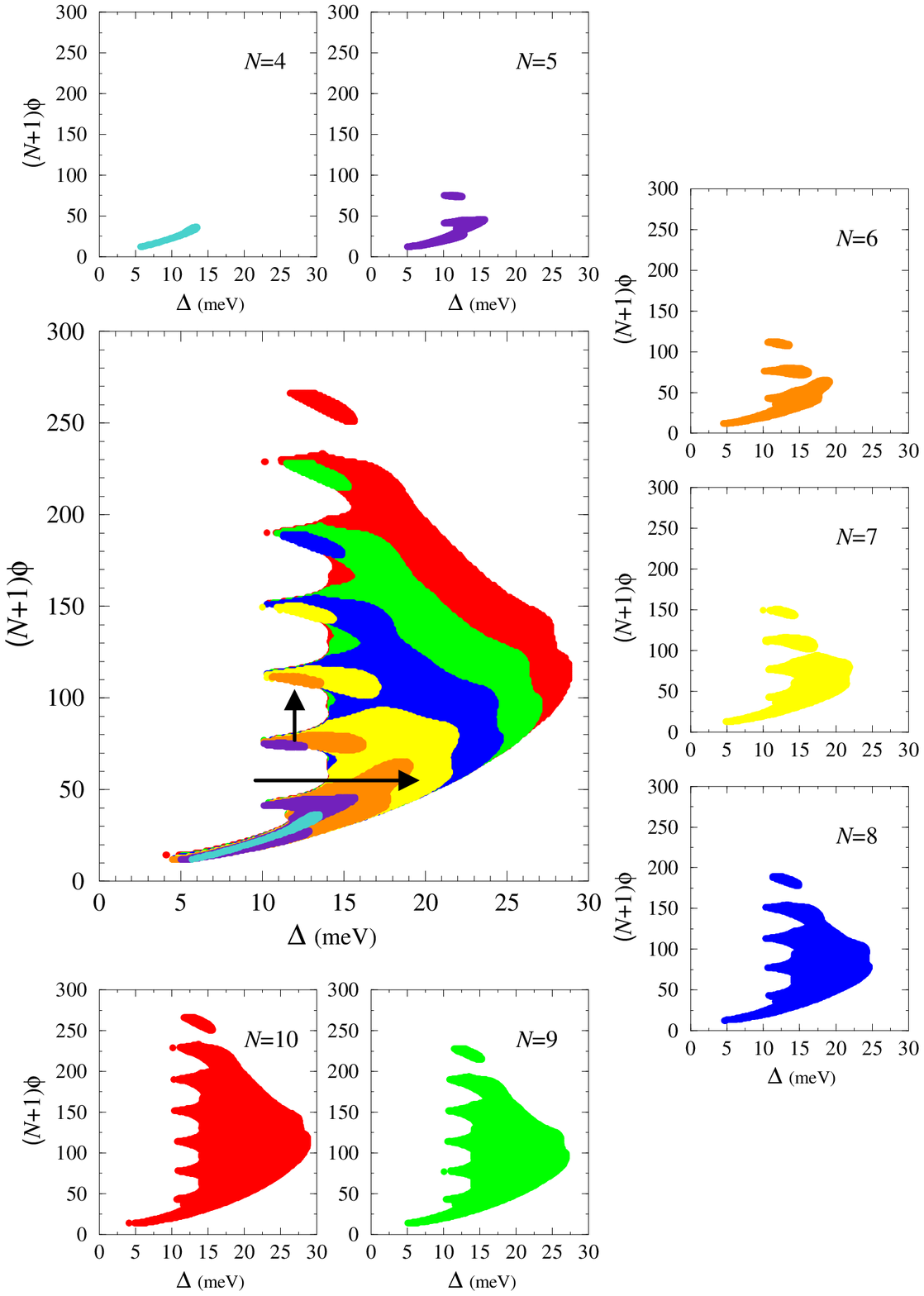}}}
\caption{(Color online) Central panel: Phase diagram of dimensionless voltage $(N+1)\phi=V/V_0$
($V_0= 0.96$ mV) versus $\Delta$ for different $N$. Lateral panels: same phase diagram but for specific
values of $N= 4,5, 6, 7, 8, 9, 10$.
The SS is stable in the white region, whereas SSCOs are stable in the shaded (colored) regions.
%Colors pale blue, violet, light brown, yellow, blue, green and red correspond to
%$N$=$4$, 5, 6, 7, 8, 9, 10.
For a given value of $N$, the SSCO regions contain those for all
smaller values of $N$.}
\label{1}
\end{figure}

\begin{figure}[ht]
\centerline{\hbox{
\epsfxsize=\textwidth
\epsfbox{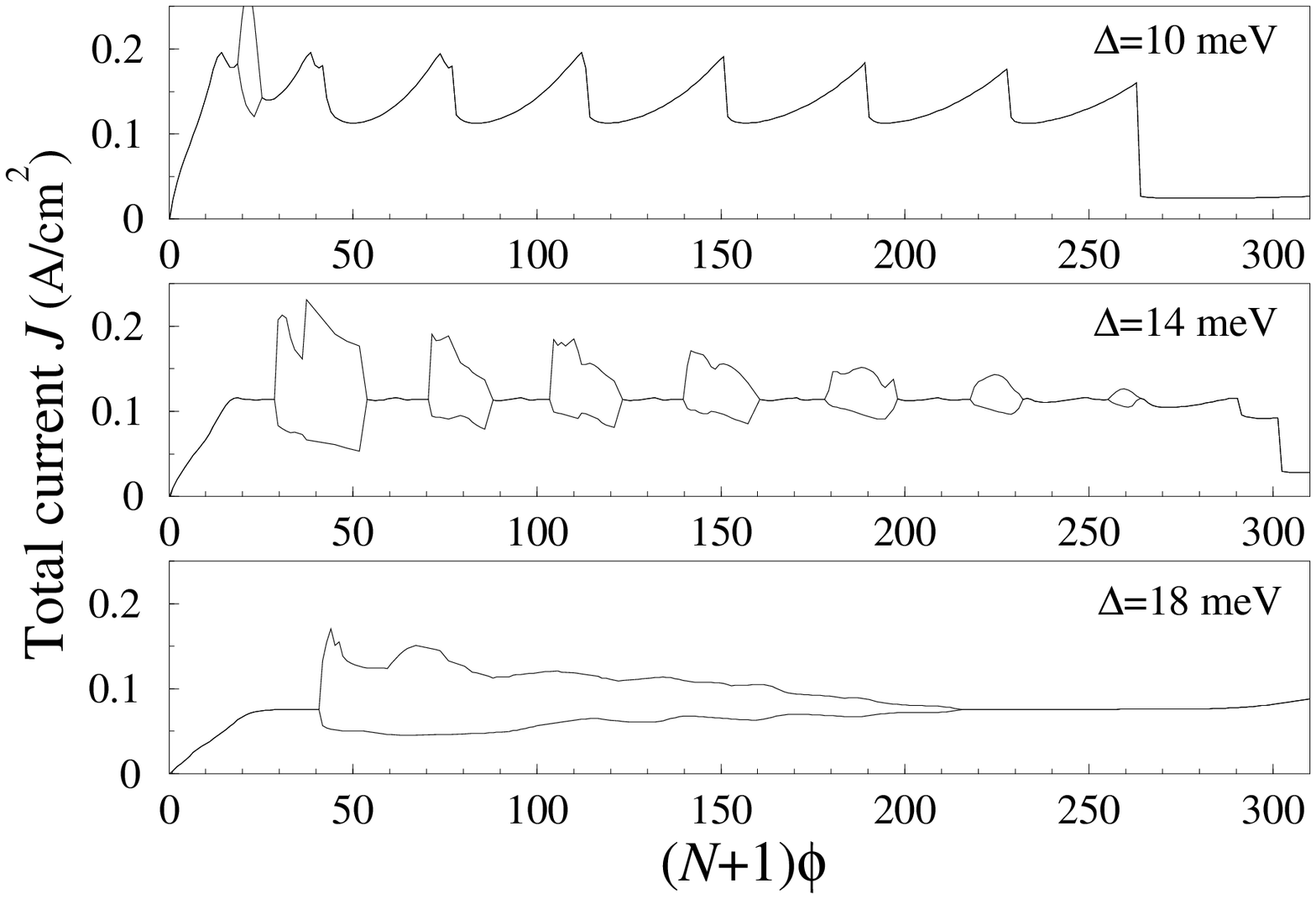}}}
\caption{$J$--$V$ characteristic curves for a MQWS with $N=10$ and level splittings of
$\Delta = 10$, 14 and 18 meV.}
\label{2}
\end{figure}

Figure \ref{1} depicts the phase diagram of voltage versus $B$-induced level splitting 
$\Delta$. We observe that the extension of the parameter regions corresponding
to SSCOs increases with $N$, the number of QWs in the structure, and that SSCO 
regions for a fixed $N$ contain SSCO regions for structures with smaller $N$. To the 
left of the main oscillatory regions there are sometimes small oscillatory regions
which appear as isolated dots in Fig.~\ref{1}. Whether these dots are connected to the
main oscillatory regions by extremely thin regions (narrower than the discretization error of the
code) is something our 
numerical solution of the model has not been able to decide.
Even though these connecting regions have not been found and therefore they are
not shown in the figure, we cannot discard their existence. In fact the dots are absorbed by the
larger oscillatory regions to their right as $N$ increases.

Figure \ref{2} shows the total current density $J(t)$ as a function of the applied
dimensionless voltage $\phi$ for $N=10$. In agreement with Fig.\ \ref{1}, we observe
that the width of SSCO regions increases with $\Delta$. For intermediate values of $\Delta$,
the number of oscillatory regions first increases and then decreases again when the oscillatory
regions merge for larger $\Delta$.

\begin{figure}[ht]
\centerline{\hbox{
\epsfxsize=\textwidth
\epsfbox{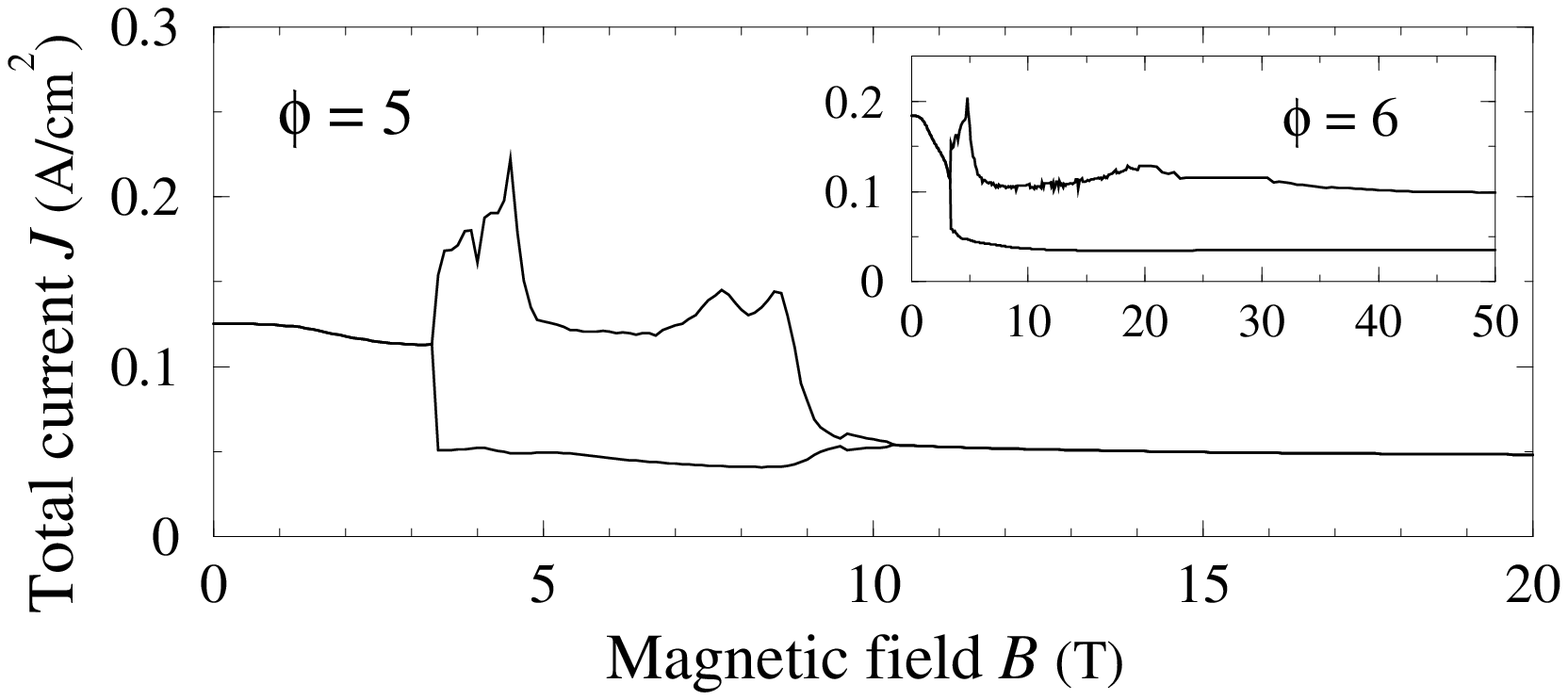}}}
\caption{$J$--$B$ plot for a MQWS with $N=10$ and $\phi$=$5$. Inset: same for $\phi = 6$,
showing indefinite persistence of SSCOs for large values of $B$.}
\label{3}
\end{figure}

The phase diagram shows similar features if we change $\Delta$ at fixed $V$: there is only
one finite interval of SSCOs, $(\Delta_{l},\Delta_{u})$, cf.\ Fig.\ \ref{1}. In terms of
the magnetic field $B$, the situation is somewhat different. The level splitting $\Delta$
is given by a Brillouin function of $B$, so $\Delta$ can only take on values smaller than
a saturating value, $\Delta_{\infty}$. If $\Delta_{u}<\Delta_{\infty}$, there may be SSCOs
for a finite interval of $B$ as shown  in Fig.~\ref{3} or for an infinite interval if
$\Delta_{\infty}<\Delta_{u}$ (inset of Fig.~\ref{3}).

\begin{figure}[ht]
\centerline{\hbox{
\epsfxsize=\textwidth
\epsfbox{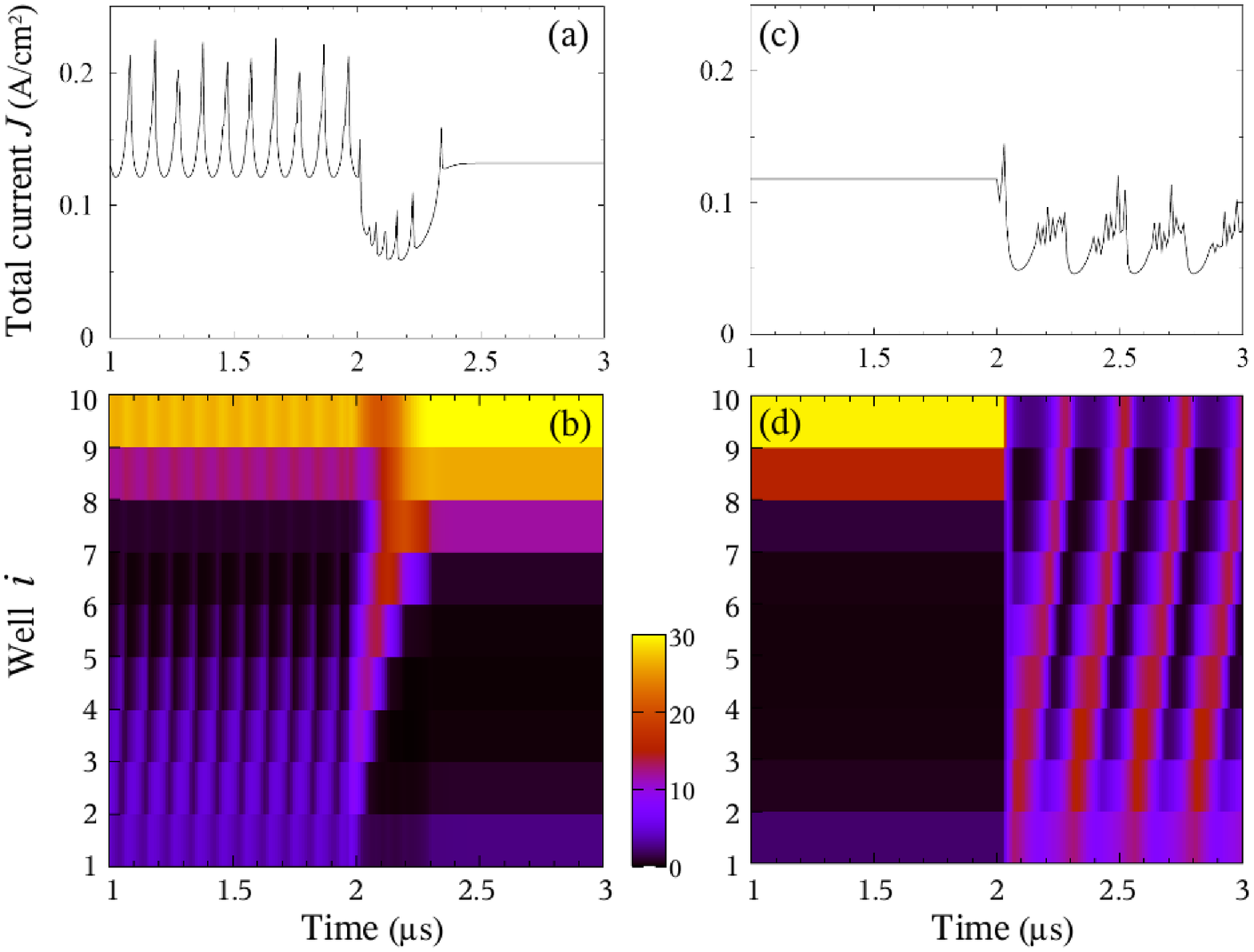}}}
\caption{(Color online) $J(t)$ curve and density plot of the electric field $F$
showing the response of a $N=10$ MQWS to the abrupt switches marked in Fig.\ \ref{1}.
(a) and (b): vertical switch from $\phi$=$7$ to 9 with $\Delta$=$12$ meV.
(c) and (d): horizontal switch from $B$=$2$ T to 6 T with $\phi$=$5$.}
\label{4}
\end{figure}

The MQWS response to a sudden switching of the voltage or the magnetic field can be
inferred from Fig.\ \ref{1}. Let us increase $\phi$ at fixed $\Delta$ as indicated by the 
vertical arrow in Fig.\ \ref{1} (from SSCOs to SSs). After a transient, the MQWS settles 
to the SS, as shown in Figs.\ \ref{4}(a) and (b).
The periodic generation of a high field domain at the magnetic QW $i$=$1$ and its motion
towards the high field region adjacent to the collector yield SSCOs (Panel b).
The transient corresponds to the nucleation of a last and larger high field domain at the
first QW and its motion until its trailing domain wall reaches the location corresponding
to the stable SS (cf.\ the similar dipole-tripole mechanism in \cite{amann,bon06,BGr05}).
Increasing abruptly $\phi$ between SSs in different regions of Fig.\ \ref{1} always involves
dipole  emission at the magnetic QW, unlike the one-well upward domain wall motion possible in 
conventional III-V MQWS \cite{amann}.

For fixed $\phi$, a sudden increment of $B$ from a stable SS region to a SSCO region (horizontal
arrow in Fig.\ \ref{1}) induces SSCOs, as shown in Fig.\ \ref{4}(c-d). The transient stage between
the SS and SSCOs after switching $B$ is due to the formation of a high field domain at the first
QW which travels towards the collector. After the domain reaches the MQWS end, a new high field
domain is formed at the first QW and the same situation is periodically repeated.

\section{Discussion}
In this paper, we have systematically analyzed the transition from stationary states to
self-sustained current oscillations through a dilute magnetic semiconductor multi-quantum
well structure. Switching suddenly a control parameter as the (dimensionless) applied voltage
$\phi$ or the external magnetic field $B$ may force the system to move between stable oscillatory and stationary states through the transition region. Since self-sustained current oscillations
are caused by triggering high field domains at the magnetic quantum well, we expect our results
not to change qualitatively with the contact boundary condition.
We have used two other conditions to check this:
\begin{enumerate}

\item[(i)] $n_{0}^\pm = n_{N+1}^\pm = \kappa N_{D}/2$ in the tunneling currents 
for normal contacts, where $\kappa$ is a positive constant.

\item[(ii)] The electric field at the injector $F_0$ is calculated by using the Ohm's law:
\begin{eqnarray}
\varepsilon\, \frac{{\rm d} F_0}{{\rm d}t} + \sigma F_0 = J(t),
\end{eqnarray}
instead of the tunneling current formulas \cite{njp} with known $n_0^\pm$.
\end{enumerate}
The resulting phase diagrams for (i) with different values of $\kappa$ (from 0.5 to 1.5) and 
for (ii) with different values of the contact resistivity $1/\sigma$ (from 31.3 to 313 
$\Omega$m) have the same configuration as in Fig.\ \ref{1} except for small quantitative 
shifts of the SSCO regions: increasing $\kappa$ or $1/\sigma$ enlarges the oscillatory 
region. 

This situation is reminiscent of the early theoretical work on the Gunn effect in bulk GaAs. 
Gunn's experiments made it clear that the SSCOs in dc voltage biased n-doped GaAs samples
are due to the periodic motion of charge dipole waves that appear at the cathode and 
disappear at the anode \cite{gunn}. Theorists soon used a variety of boundary conditions at the injecting 
contact region (cathode) that could produce the required SSCOs mediated by charge dipole 
waves. Among them, Kroemer's contact characteristics (the electron current density is a 
known function of the electric field at the contact: in case (ii), this function is linear) 
\cite{kro68}, general models of metal--semiconductor contacts \cite{gom97}, or contacts
with fixed electron density but with a notch near the cathode in the doping density profile \cite{mcc66}. 
While the two first contact types can give rise to SSCOs mediated by either moving charge 
dipoles (either high or low electric field domains) or monopoles (either charge accumulation
or charge depletion layers) depending on parameter values \cite{hig92,gom97}, a notch in 
the doping density produces only dipoles \cite{kro68,mcc66}.

While it is feasible to list all 
possible oscillation types in terms of contact parameter values (see \cite{gom97,hig92}), 
these values cannot be modified once the Gunn diode has been made. Similarly, in a 
conventional III-V weakly coupled n-doped semiconductor superlattice (SL), the boundary condition at 
the injector, the SL configuration and the doping density at the QWs determine whether the 
system exhibits SSCOs mediated by charge dipole or monopole waves \cite{san99,BGr05} 
or multistable static electric field domains; see the review \cite{BGr05} and references cited 
therein.

In the SL case, there exist partial phase diagrams: (i) doping density vs dc voltage 
bias for fixed boundary condition \cite{MGB00}, and (ii) injector conductivity vs dc 
voltage bias for fixed doping density (assuming a linear relation between electron current 
and electric field at the contact) \cite{jilis}. However a complete study (which, depending
on both doping density {\em and} injector conductivity, should yield 
both monopole {\em and} dipole SSCOs, as in \cite{san99}) has not yet been carried out. 
Be this as it may, once the SL has been made and contacted, the stable solutions can be 
selected only by changing the bias and this limits the type of attractors present in a
particular SL.

The situation is different in the case of a dilute magnetic semiconductor multi-quantum
well structure: the magnetic QW plays the role of a ``tunable doping density notch''.
In principle, any self-oscillations that may appear are due to triggering of dipoles at
the magnetic QW. However, by changing the external magnetic field we can select either
stable stationary states or SSCOs as the DMS multi-quantum well response.

Our results show how to design a device operating a spin injector and a spin oscillator
by tuning the Zeeman splitting and the parameters determining the sample configuration. 

%\ack
\section{Acknowledgment}
Work supported by the MICINN grants FIS2008-04921-C02-01 and
MAT2008-02626. R.E. thanks the Spanish Ministry of Education Ram\'on y Cajal Program.

%\newpage
%\section*{References}

\end{document}